\documentstyle[preprint,aps]{revtex}
\begin{document}

\draft
\preprint{LBL-37539}

\title{Partial $U(1)_A$ Restoration and $\eta$ Enhancement\\
 in High-Energy Heavy-Ion Collisions
%\thanks{\baselineskip=12pt
%This work was
%supported
%by the Director, Office of Energy
%Research, Office of High Energy and Nuclear Physics, Divisions of High
%Energy Physics and Nuclear Physics
% of the U.S. Department of Energy under Contract
%DE-AC03-76SF00098, and by the Natural Sciences and
%Engineering Research Council of Canada.}
    }
\author{Zheng Huang$^a$ \thanks{Address after August 1, 1995:
Department of Physics, University of Arizona,
Tucson, AZ 85721.}
and Xin-Nian Wang$^b$}

\address{{
$^a$Theoretical Physics Group and
$^b$Nuclear Science Division\\
Lawrence Berkeley Laboratory,
University of California,
    Berkeley, CA 94720, USA}}

\date{July 12, 1995}
\maketitle
\begin{abstract}
We calculate the thermally averaged rates for the
 $\eta$-$\pi$ conversion and $\eta$ scattering
using the Di Vecchia-Veneziano model and t' Hooft model, which
incorporate explicitly the $U(1)_A$ anomaly.
Assuming an exponential suppression of the $U(1)_A$ anomaly, we also
take into account the partial restoration of $U(1)_A$
symmetry at high temperatures. We find
that the chemical equilibrium between $\eta$ and $\pi$ breaks
up considerably earlier than the thermal equilibrium.
Two distinct scenarios for the $\eta$ freeze-out are discussed
and the corresponding  chemical potentials are calculated.
We predict an enhancement of the thermal $\eta$-production
as a possible signal of the partial $U(1)_A$ restoration in
high-energy heavy-ion collisions.
\end{abstract}

\pacs{11.30.Rd, 11.30.Qc, 12.38.Mh}

\newpage
\narrowtext

\section{Introduction}

At the Lagrangian level, QCD has, in addition to $SU(N_f)\times SU(N_f)$
chiral symmetry, an approximate $U(1)_A$ symmetry, under which
all left-handed quark fields are rotated by a common phase while the
right-handed quark fields are rotated by an opposite phase. It is well
known that the $U(1)_A$ symmetry is violated by the axial anomaly present at
the quantum level and thus cannot give rise to the Goldstone boson
which would occur when $U(N_f)\times U(N_f)$ chiral symmetry is
spontaneously broken. The $U(1)_A$ particle, known as $\eta '(958)$
in the $N_f=3$ case,  acquires an additional mass
through the quantum tunneling effects mediated by instantons \cite{thoo},
breaking up the mass degeneracy with pions and $\eta$ in the chiral limit
when all quarks ($u$, $d$ and $s$) are massless. The $\eta (547)$
particle also acquires an additional mass through the mixing with
$\eta '$. It is believed that at high temperatures the instanton
effects are suppressed due to the Debye-type screening \cite{gpy}. Then one
expects a practical restoration of $U(1)_A$ at high temperatures. If the
restoration occurs at a temperature lower than the chiral phase transition
temperature $T_{\chi}$, there may be some interesting phenomenological
implications in high-energy heavy-ion collisions, as suggested first
by Pisarski and Wilczek \cite{pisa} and more recently by Shuryak \cite{shur}.
One of the consequences of $U(1)_A$ restoration is the enhancement of
$\eta$ particle production at small and intermediate transverse momenta
due to the softening of its mass at high temperatures. However, the
final yield of the $\eta$ particles and their $p_t$ distributions both
depend crucially on the chemical and thermal equilibrating processes
involving the $\eta$.

In this paper, we shall examine the rates of various processes relevant
for the thermal $\eta$ particle production, in particular,
 whether or not the $\eta$ can decouple early enough from the thermal
system expected to be produced in relativistic heavy ion collisions. We
shall present a theoretical calculation of the thermal cross sections
for the processes $\eta\eta \leftrightarrow \eta\eta$,
$\pi\eta \leftrightarrow \pi\eta$
and $\eta\eta \leftrightarrow \pi\pi$, essential
to the thermal and chemical equilibration.
Our calculations are based on models which explicitly incorporate
the $U(1)_A$ anomaly. We also assume an exponential suppression of
the $U(1)_A$ anomaly due to the Debye-type screening of the instanton
effect \cite{gpy}, which leads to the temperature dependence of the $\eta$
and $\eta '$ masses. Our results
suggest that the chemical equilibrium breaks up for $\eta$ particles long
before the thermal freeze-out. We suggest a modest enhancement of
thermal  $\eta$ production as a signal for the relic of $U(1)_A$ restoration.

This paper is organized as follows: In Sec.\ \ref{II}, we compute the mass
spectrum of $\eta$ and $\eta '$ using the Di Vecchia-Veneziano model,
which incorporates the $U(1)_A$ anomaly and the $\eta$-$\eta '$
mixing effect. We obtain the low-energy theorems for various scattering
amplitudes. In Sec.\ \ref{III}, we incorporate the $\sigma$ and the $\delta$
resonances using the t' Hooft model and reevaluate the $\eta$ scattering
cross sections. In Sec.\ \ref{IV}, we study the thermal averaged
cross sections responsible for maintaining thermal and chemical
equilibria, and suggest that the chemical equilibrium between
$\eta$ and $\pi$ breaks
up considerably earlier than the $\eta$ thermal equilibrium.
We discuss two scenarios for the $\eta$ freeze-out and their corresponding
signals for the $\eta$ production.
We briefly comment on the roles of $\eta '$ and the QCD sphalerons in
Sec.\ \ref{V} and Sec.\ \ref{VI} respectively.

\section{Nonlinear $\sigma$-Model: Low-energy Theorems}
\label{II}
Up to now, there has been no direct experimental measurement of the
$\eta$ scattering cross sections (or the scattering lengths). One has to
rely on theoretical models to calculate the interaction rates which
are complicated by many uncertainties. Nevertheless, the scattering
amplitudes at low energy can be more or less precisely predicted if
the meson masses are soft, thanks to the soft-meson theorems which
are based on the symmetry of the interactions and
depend very little on the detailed dynamics. The current-algebra predictions
of these scattering amplitudes have been made very early by Osborn \cite{osbo}
based on $SU(3)\times SU(3)$, where the anomalous $U(1)_A$ and the $\eta$
and $\eta '$ mixing are not included. In the light of softening of
$\eta$ and $\eta '$ masses at high temperatures, we argue that the symmetry
can be extended to $U(3)\times U(3)$. We shall rederive the low-energy
amplitudes incorporating the anomalous $U(1)_A$ using the
nonlinear $\sigma$-model that at the lowest order should give us the low-energy
theorems. The standard Di Vecchia-Veneziano model \cite{vene,chen},
which incorporates the explicit $U(1)_A$ anomaly, reads after
integrating out the gluon field
\begin{equation}
{\cal L}_{\rm eff}=\frac{f_\pi^2}{4}{\rm Tr} (\partial^\mu U
\partial_\mu U^\dagger )+ \frac{f_\pi^2}{4}{\rm Tr}(MU+MU^\dagger )
+\frac{f_\pi^2}{4}\frac{a}{4N_c}(\log {\rm det}U-\log {\rm det}U^\dagger )^2,
\label{1}
\end{equation}
where $U=\exp (i\Phi/f_\pi)$, $f_\pi=93$ MeV,
$M={\rm diag}(m^2_\pi,m^2_\pi,2m_K^2-m^2_\pi )$ and
\begin{equation}
\Phi=\left(
\begin{array}{ccc}
\pi^0+\eta_8/\sqrt{3}+\sqrt{2}\eta_1/\sqrt{3} &
\sqrt{2}\pi^+ & \sqrt{2}K^+\\
\sqrt{2}\pi^- & -\pi^0+\eta_8/\sqrt{3}+\sqrt{2}
\eta_1/\sqrt{3} & \sqrt{2}K^0\\
\sqrt{2}K^- & \sqrt{2}\overline{K}^0 & -2\eta_8/\sqrt{3}+
\sqrt{2}\eta_1/\sqrt{3}
\end{array}
\right) \; .
\end{equation}
The last term in
Eq.~(\ref{1}) is the anomaly term which breaks $U(1)_A$ explicitly. It is
easy to check that Eq.~(\ref{1}) satisfies the anomalous Ward identity which
is crucial for determining the form of $U(1)_A$ breaking \cite{hatsuda}.
In Eq.~(\ref{1}), $a$ is related to the topological charge correlation
function in pure  Yang-Mills theory
\begin{equation}
a=-i\frac{6}{f_\pi^2}\int d^4x \langle T[F_{\mu\nu}\tilde{F}^{\mu\nu}(x)
F_{\mu\nu}\tilde{F}^{\mu\nu}(0)]\rangle_{\rm YM}\; ,
\label{3}
\end{equation}
where $\tilde{F}^{\mu\nu}$ is the dual gluon field strength tensor and
$\langle \cdots \rangle$ stands for the vacuum expectation value at zero
temperature or the thermal average at finite temperature.
The integral $a$ is identically
zero in perturbation theory; it only receives  nonperturbative contributions
arising from the topologically nontrivial instanton configurations. The
calculation of $a$ at both zero and finite temperature is done by Gross,
Pisarski and Yaffe \cite{gpy} using a dilute gas approximation, and by
Dyakonov and Petrov and by Shuryak
 \cite{dyak} using an instanton liquid model. For our
purpose, the phenomenological value of $a$ at $T=0$ can be fixed by the meson
mass spectroscopy, while $a(T\neq 0)$ will be modeled by assuming an
exponential suppression shown by Pisarski and Yaffe \cite{yaff} at
high $T$.

The quadratic terms for the octet $\eta_8$ and the singlet $\eta_1$
from the Lagrangian reads
\begin{equation}
{\cal L}_{\rm mass}=-\frac{1}{2}\left [
\left( -\frac{m_\pi^2}{3}+\frac{4m_K^2}{3}\right)\eta_8^2+
\left(\frac{2m_K^2}{3}+
\frac{m_\pi^2}{3}+a\right)\eta_1^2+\frac{2\sqrt{2}}{3}\left(
2m_\pi^2-2m_K^2\right)\eta_8
\eta_1\right ] \; .
\label{4}
\end{equation}
Clearly, there is a mixing between the octet $\eta_8$ and the
singlet $\eta_1$.
The physical $\eta (547)$ and $\eta '(958)$ are defined by
\begin{equation}
\eta =\eta_8\cos \theta +\eta_1\sin\theta\quad\quad ; \quad\quad
\eta '=-\eta_8\sin\theta +\eta_1\cos\theta
\end{equation}
to diagonalize the quadratic terms with the mixing angle
\begin{equation}
\tan \theta =\frac{4m_K^2-m_\pi^2-3m_\eta^2}{2\sqrt{2}(m_K^2-m_\pi^2)}\; ,
\label{5}
\end{equation}
and the physical masses are
\begin{eqnarray}
m_\eta^2 & = & (m_K^2+a/2)-\frac{1}{2}\sqrt{(2m_K^2-2m_\pi^2-a/3)^2+8a^2/9}
 \; , \label{6} \\
m_{\eta '} ^2 & = & (m_K^2+a/2)+
\frac{1}{2}\sqrt{(2m_K^2-2m_\pi^2-a/3)^2+8a^2/9} \; .\label{6b}
\end{eqnarray}
The mixing angle $\theta$ as well as $m_\eta^2$ and $m_{\eta '}^2$
depend on the instanton-induced quantity $a$ which is a
function of temperature.
The precise form of $a(T)$ at low temperature is not known. Nevertheless,
if the $U(1)_A$ breaking becomes soft at a temperature lower than
the chiral phase transition temperature $T_{\chi}$,
one may model the suppression effect by an exponential dependence
\cite{yaff,kiku,kuni}
\begin{equation}
a(T)=a(0)e^{-(T/T_0)^2} \; , \label{7}
\end{equation}
where $T_0\simeq 100 - 200$ MeV, while keeping the masses of the pion and kaon
approximately temperature independent, since they change very slowly with
the temperature. It is known that mixing angle $\theta$,
$m_\eta^2$ and $m_{\eta '}^2$ at $T=0$ cannot be simultaneously fit to their
experimental values by a single parameter $a(0)$. The best fit is to use
the measured value of $m_\eta^2+m_{\eta '}^2$ as an input to determine
$a(0)=(m_\eta^2+m_{\eta '}^2)-2m_K^2$ and use this  $a(0)$ to predict
$\theta$, $m_\eta^2$ and $m_{\eta '}^2$  using Eqs.~(\ref{5}) and
(\ref{6}). At $T=0$, the predicted values are $\theta =18.3^0$,
$m_\eta = 500$ MeV and $m_{\eta '}=984$ MeV, compared to the measured
values $\theta^{\rm exp} \simeq 20^0$ from
$\eta ,\eta '\rightarrow \gamma\gamma$,
$m_\eta^{\rm exp}=547$ MeV and $m_{\eta '}^{\rm exp}=958$ MeV.
The temperature dependence of $m_\eta$ and $m_{\eta '}$
is completely determined by a temperature-dependent $a(T)$
given in Eq.~(\ref{7}).
Throughout this paper, we take $T_0=150$ MeV in Eq.~(\ref{7}).
It should be emphasized  that in relativistic heavy-ion collisions,
the thermal system freezes out at about $T_{\rm th}=130 -  150$ MeV,
when the collision time scale exceeds the size of the system
mainly determined by the nuclear radius $R= 4 - 8$ fm for central
$S+S$ or $Pb+Pb$ collisions. Below the freeze-out temperature $T_{\rm th}$,
 the finite-temperature calculation of $a(T)$ does not make sense
and the behavior of $a$ is determined by the nonequilibrium dynamics.
Figure 1 schematically
plots such a temperature dependence. Clearly, the $\eta$ becomes
soft at high $T$ and eventually is degenerate with the pions.
The mass of $\eta '$ also decreases. However, it does
not become degenerate with the pion because of the large strange-quark
mass, as is seen from Eq.~(\ref{6b}). From Fig.~1 we see that the $\eta '$
mass at high temperatures is still higher than the $\eta$ mass
at zero temperature. In Fig.~1 we also plot the temperature
dependence of the $\delta$ resonance mass which we will discuss in
the following section. At temperatures higher than $T_\chi$, the masses of
these excitation modes will all rise again.

The interaction terms are obtained by expanding $U$ in Eq.~(\ref{1}).
In contrast to the pion field, there are no derivative
couplings involving $\eta$ and $\eta '$.
We shall ignore the interactions of $\eta$ and $\eta '$
with kaons since they are heavy compared with pions. To the lowest order, the
quartic terms involving $\pi$, $\eta$ and $\eta '$ are
\begin{eqnarray}
{\cal L}_{\rm int} & = &\frac{1}{2\times 4!f_\pi^2}\left[
2m_\pi^2(\eta \sin \chi +\eta '\cos\chi )^4+(4m_K^2-2m_\pi^2)
(\eta '\sin\chi -\eta\cos\chi )^4\right. \nonumber\\
& & \left. +12m_\pi^2
\mbox{\boldmath $\pi$}^2(\eta\sin\chi +\eta '\cos\chi )^2 \right]\; ,
\label{8}
\end{eqnarray}
where $\chi =\theta +\arctan 1/\sqrt{2}$.  At very high $T$, as
$\theta \rightarrow \arctan \sqrt{2}$ and $\chi\rightarrow \pi/2$,
we can see that the $\eta '$ decouples from interactions with $\pi$
and $\eta$. The low-energy theorems on the two-body scattering
amplitudes can be easily derived from Eq.~(\ref{8}):
\begin{eqnarray}
{\cal A}(\eta\eta\leftrightarrow\eta\eta ) & = &
\frac{1}{f_\pi^2}[m_\pi^2\sin ^4\chi
+(4m_K^2-2m_\pi^2)\cos ^4\chi ] \; ,\nonumber\\
{\cal A}(\eta\eta\leftrightarrow\pi^a\pi^a ) & = & {\cal A}(\pi^a\eta
\leftrightarrow \pi^a\eta )=
\frac{1}{f_\pi^2}m_\pi^2\sin ^2\chi \; ,\nonumber\\
{\cal A}(\eta '\eta '\leftrightarrow\pi^a\pi^a ) & = & {\cal A}(\pi^a\eta '
\leftrightarrow \pi^a\eta ')=
\frac{1}{f_\pi^2}m_\pi^2\cos ^2\chi \; ,\nonumber\\
{\cal A}(\eta\eta '\leftrightarrow\pi^a\pi^a ) & = & {\cal A}(\pi^a\eta '
\leftrightarrow \pi^a\eta )=
\frac{1}{f_\pi^2}m_\pi^2\sin \chi\cos\chi \; ,\nonumber\\
{\cal A}(\eta '\eta '\leftrightarrow\eta '\eta ' ) & = &
\frac{1}{f_\pi^2}[m_\pi^2\cos ^4\chi
+(4m_K^2-2m_\pi^2)\sin ^4\chi ]  \; ,\nonumber\\
{\cal A}(\eta '\eta '\leftrightarrow\eta \eta  ) & = & {\cal A}(\eta\eta '
\leftrightarrow \eta\eta '=
\frac{1}{f_\pi^2}(4m_K^2-m_\pi^2)\sin ^2\chi\cos^2\chi \; ,\nonumber\\
{\cal A}(\eta \eta \leftrightarrow\eta \eta ' ) & = &
\frac{1}{f_\pi^2}[m_\pi^2\sin^3\chi\cos \chi
-(4m_K^2-2m_\pi^2)\sin\chi\cos^3\chi ]\; , \nonumber\\
{\cal A}(\eta \eta ' \leftrightarrow\eta ' \eta ' ) & = &
\frac{1}{f_\pi^2}[m_\pi^2\sin\chi\cos^3 \chi
-(4m_K^2-2m_\pi^2)\sin^3\chi\cos\chi ]\; . \label{9}
\end{eqnarray}
The results calculated by Osborn \cite{osbo} based on the current algebra can
be recovered by taking $\theta =0$ and using the Gell-Mann-Okubo relation
$4m_K^2=3m_\eta^2+m_\pi^2$. These low-energy theorems must be satisfied by any
dynamical models, because they are solely based on
the symmetry properties of the theory.

\section{Linear $\sigma$-Model: Inclusion of Resonances}
\label{III}
The amplitudes listed in Eq.~(\ref{9}) grossly underestimate the strength of
scatterings at higher energies, especially in the resonance regions.
However, the inclusion of resonances introduces many uncertainties,
such as which resonances should be included and what are the couplings
of these resonances to the mesons. In addition, there is no
guarantee that a naive lowest-order calculation will preserve the
unitarity because of the strong interactions.
Fortunately, the low-energy theorems provide us some guidelines as to how
the amplitudes should approach their low-energy limits. The linear
$\sigma$-model based on the chiral symmetry is known to satisfy the low-energy
theorems, and at the same time to be able to incorporate the resonances.
To further reduce the input parameters, we consider the $\sigma$ and
$\delta (980)$ [now called $a_0(980)]$ resonances, which, together with
$\pi$ and the $\eta_{\rm ns}$ to be defined below, form a complete
representation of $U(2)\times U(2)$. We shall concentrate on the $\eta$
particle, since there is no dramatic change of the $\eta '$ mass
with temperature, as shown in Fig.\ 1.
We study the most relevant processes for the $\eta$ production:
$\eta\eta\leftrightarrow \pi\pi$,
$\pi\eta\leftrightarrow \pi\eta$, and
$\eta\eta\leftrightarrow \eta\eta$. In this case, $U(3)\times U(3)$ reduces
to $U(2)\times U(2)$ except for the mixing effects which we have already
calculated.

Let us introduce the nonstrange
mode $\eta_{\rm ns}=(u\bar u+d\bar d)/\sqrt{2}$ and
take $m_s$ to be heavy. Then $\eta_{\rm ns}$ is approximately a mass
eigenstate, $\eta_{\rm ns}=\eta \sin\chi +\eta '\cos\chi $, whose mass is
determined from Eq.~(\ref{4}) to be $m_{\rm ns}^2\simeq 2a/3+m_\pi^2$. At zero
temperature, $m_{\rm ns}\simeq 709$ MeV. We then define the (2,2)
representation multiplet of $U(2)\times U(2)$ as
\begin{equation}
\Phi =\frac{1}{2}(\sigma +i\eta_{\rm ns})+\frac{1}{2}(\mbox{\boldmath
$\delta$} +i\mbox{\boldmath
$\pi$})\cdot \mbox{\boldmath
$\tau$}\; .
\end{equation}
The most general $U(2)\times U(2)$ invariant potential is
\begin{equation}
V_0=-\mu^2{\rm Tr}(\Phi^\dagger\Phi )+\frac{1}{2}(\lambda_1-\lambda_2)
({\rm Tr}\Phi^\dagger\Phi )^2+\lambda_2{\rm Tr}(\Phi^\dagger\Phi )^2
\label{10}
\end{equation}
and the mass term is
\begin{equation}
V_m=\frac{m_\pi^2f_\pi}{4}{\rm Tr}(\Phi^\dagger +\Phi )\; ,\label{11}
\end{equation}
where $\lambda_1$, $\lambda_2$ are dimensionless constants.
The $U(1)_A$-breaking term, consistent with the Ward identity,
is introduced by t' Hooft \cite{thoo2} as
\begin{equation}
V_a=\frac{a}{3}({\rm det}\Phi^\dagger +{\rm det}\Phi )\; ,
\label{12}
\end{equation}
and the coefficient
in $V_a$ is chosen such that it gives the correct mass for $\eta_{\rm ns}$.
The mass spectrum can be derived from Eqs.~(\ref{10}), (\ref{11}) and
(\ref{12}) by making a shift $\sigma\rightarrow f_\pi +\sigma$:
\begin{equation}
m_\sigma^2=\lambda_1f_\pi^2+m_\pi^2\quad ;\quad
m_\delta^2=\lambda_2f_\pi^2+m_{\rm ns}^2 \; .\label{13}
\end{equation}
The decay widths are
\begin{eqnarray}
\Gamma_\sigma & = & \frac{3}{32\pi}(m_\sigma^2-4m_\pi^2)^{1/2}
\frac{(m_\sigma^2-m_\pi^2)^2}{f_\pi^2m_\sigma^2} \; ,\\
\Gamma_\delta  & = & \{[(m_\delta^2-(m_{\rm ns}+m_\pi )^2]
[(m_\delta^2-(m_{\rm ns}-m_\pi )^2]\}^{1/2}
\frac{(m_\delta^2-m_{\rm ns}^2)^2}{16\pi f_\pi^2m_\delta^3}\; .
\end{eqnarray}
At zero temperature, $\Gamma_\sigma \sim 1$ GeV (if $m_\sigma \sim 700$ MeV)
and $\Gamma_\delta \sim 200$ MeV.
In principle, we should also take into account the temperature
dependence of $f_\pi$ and $m_\sigma$ below $T_\chi$.
Here, we assume the chiral phase transition is very rapid after which
$f_\pi$ and $m_\sigma$ have very slow temperature dependences.
Furthermore, due to the large width of the $\sigma$, the slow
temperature dependence of $m_\sigma$ will not change our
results significantly. Under such an assumption, the linear $\sigma$-model
predicts also some softening of the $\delta$ resonance as
$T$ increases, because $\delta$ is the chiral partner of $\pi$ and acquires
some mass from the $U(1)_A$ anomaly. The temperature dependence of $m_\delta$
 is plotted in Fig.\ 1.

The interaction terms are
\begin{eqnarray}
{\cal L}_{\rm int} & = & \frac{\lambda_1f_\pi}{2}(\sigma^2+\eta_{\rm ns}^2
+\mbox{\boldmath $\delta$}^2+\mbox{\boldmath $\pi$}^2)\sigma
+\frac{\lambda_1}{8}(\sigma^2+\eta_{\rm ns}^2+\mbox{\boldmath $\delta$}^2
+\mbox{\boldmath $\pi$}^2)^2 \nonumber \\
& & +\lambda_2f_\pi(\sigma\mbox{\boldmath $\delta$}+
\eta_{\rm ns}\mbox{\boldmath $\pi$})\cdot \mbox{\boldmath $\delta$}
+\frac{\lambda_2}{2}(\sigma\mbox{\boldmath $\delta$}+
\eta_{\rm ns}\mbox{\boldmath $\pi$})^2+\frac{\lambda_2}{2}
(\mbox{\boldmath $\delta$}\times \mbox{\boldmath $\pi$})^2\; .\label{15}
\end{eqnarray}
The coupling constants $\lambda_1$ and $\lambda_2$ can be obtained
from the mass relations of Eq.~(\ref{13}). It is worth pointing out
that the above model should not be used to estimate the pion-pion
scattering amplitude, because it does not include the important
vector resonances such as $\rho$ and $A_1$. However,
since $\eta\eta$ and $\pi\eta$ scatterings cannot go
through $J=1$ channel, they do not directly affect the
interaction rates for $\eta$. Similarly, we have also neglected
the $\eta$-$\rho$ interaction.

To calculate the scattering amplitudes at the lowest order,
we have to remove a pole singularity encountered
when a resonance appears in the $s$-channel. A naive
introduction of Breit-Wigner resonance width will spoil the delicate
cancellation between the contact interaction and the pole exchange at
low energy, leading to the violation of the low-energy theorems. We adopt
a minimal prescription to save the low-energy limit developed by
Chanowitz and Gaillard \cite{chan},  making the following replacement
\begin{equation}
\lambda_1+\frac{\lambda_1^2f_\pi^2}{s-m_\sigma^2+im_\sigma\Gamma_\sigma }
\rightarrow \lambda_1(1-i\Gamma_\sigma /m_\sigma )
\frac{s-m_\pi^2}{s-m_\sigma^2+im_\sigma\Gamma_\sigma }\; .\label{16}
\end{equation}
The scattering amplitudes are calculated as follows:
\begin{eqnarray}
{\cal A}(\eta\eta\leftrightarrow \eta\eta) & = &
\sin^4\chi {\cal A}(\eta_{\rm ns}\eta_{\rm ns}
\leftrightarrow \eta_{\rm ns}\eta_{\rm ns})\nonumber\\
&= & \sin^4\chi\lambda_1(1-i\Gamma_\sigma/m_\sigma )\left[
\frac{s-m_\pi^2}{s-m_\sigma^2+im_\sigma\Gamma_\sigma }\right. \nonumber \\
& &\left. +\frac{t-m_\pi^2}{t-m_\sigma^2+im_\sigma\Gamma_\sigma }
+\frac{u-m_\pi^2}{u-m_\sigma^2+im_\sigma\Gamma_\sigma }\right] \: ,\nonumber\\
{\cal A}(\eta\eta\leftrightarrow \pi^a\pi^a) & = &
\sin^2\chi {\cal A}(\eta_{\rm ns}\eta_{\rm ns}
\leftrightarrow \pi^a\pi^a)\nonumber\\
&= & \sin^2\chi\lambda_1(1-i\Gamma_\sigma/m_\sigma )
\frac{s-m_\pi^2}{s-m_\sigma^2+im_\sigma\Gamma_\sigma }\nonumber \\
& & +\sin^2\chi\lambda_2(1-i\Gamma_\delta/m_\delta ) \left[
\frac{t-m_{\rm ns}^2}{t-m_\delta^2+im_\delta\Gamma_\delta }
+\frac{u-m_{\rm ns}^2}{u-m_\delta^2+im_\delta\Gamma_\delta }\right]
       \: ,\nonumber\\
{\cal A}(\eta\pi^a\leftrightarrow \eta\pi^a) & = &
\sin^2\chi {\cal A}(\eta_{\rm ns}\pi^a
\leftrightarrow \eta_{\rm ns}\pi^a)\nonumber\\
&= & \sin^2\chi\lambda_1(1-i\Gamma_\sigma/m_\sigma )
\frac{t-m_\pi^2}{t-m_\sigma^2+im_\sigma\Gamma_\sigma }\nonumber \\
& &+\sin^2\chi\lambda_2(1-i\Gamma_\delta/m_\delta ) \left[
\frac{s-m_{\rm ns}^2}{s-m_\delta^2+im_\delta\Gamma_\delta }
+\frac{u-m_{\rm ns}^2}{u-m_\delta^2+im_\delta\Gamma_\delta }\right] \; .
\end{eqnarray}
The cross sections for these processes are readily calculated
by integrating out
the scattering angle in $u$ and $t$, most conveniently in the CM frame:
\begin{equation}
\sigma =\frac{f}{32\pi s}\frac{|{\bf p}_{3{\rm cm}}|}{|{\bf p}_{1{\rm cm}}|}
\int^1_{-1}|{\cal A}|^2d\cos\theta \; ,\label{18}
\end{equation}
where $f=(1)1/2$ for (non-)identical particles in the final state.

\section{Thermal Production of the $\eta$ Particle}
\label{IV}
We are interested in the production of $\eta$ from a thermal source. To
learn about the thermal history of the $\eta$,
one needs to calculate the thermal
averaged cross sections for various reaction channels. Since we are only
concerned with the qualitative picture, we assume throughout the rest of
this paper Boltzmann distribution functions for thermalized $\pi$'s and
$\eta$'s and ignore the quantum Bose-Einstein enhancement. The thermal
averaged cross section for $i+j\rightarrow k+l$ is
\begin{equation}
\langle v_{ij}\sigma_{ij}(T)\rangle =
\frac{1}{8T}\frac{\int_{\sqrt{s_0}}^\infty d\sqrt{s}\sigma_{ij}(\sqrt{s})
\lambda (s,m_i,m_j )K_1(\sqrt{s}/T)}{m_i^2m_j^2K_2(m_i/T)
K_2(m_j/T)} \; , \label{19}
\end{equation}
where $\lambda (s,m_i,m_j)=[s-(m_i+m_j)^2][s-(m_i-m_j)^2]$ and $\sqrt{s_0}$
is the reaction threshold. The reactions $\eta\eta\rightarrow \eta\eta$ and
$\pi\eta\rightarrow \pi\eta$  determine
the collision time scale responsible for maintaining the thermal equilibrium
while $\eta\eta\rightarrow \pi\pi$ is responsible for the chemical equilibrium
between $\pi$'s and $\eta$'s. We define the time scales $\tau_{\rm ther}$
and $\tau_{\rm chem}$ as
\begin{eqnarray}
\tau^{-1}_{\rm ther} & = &
\langle v\sigma (\eta\eta \rightarrow \eta\eta )\rangle n_\eta +
\langle v\sigma (\eta\eta \rightarrow \pi\pi )\rangle n_\eta +
\langle v\sigma (\pi\eta \rightarrow \pi\eta )\rangle n_\pi \; , \nonumber \\
\tau^{-1}_{\rm chem} & = &
\langle v\sigma (\eta\eta \rightarrow \pi\pi )\rangle n_\eta\; ,\label{20}
\end{eqnarray}
respectively, where $n_\pi$ and $n_\eta$ are the number
densities for $\pi$ and $\eta$,
and the summation over different pion states is understood. We have performed
a numerical integration in Eq.~(\ref{19}) and plotted $\tau_{\rm ther}$
and $\tau_{\rm chem}$ as functions of the temperature in Figure 2.
In the calculation, we have explicitly taken into account the temperature
dependence of $m_\eta (T)$, $m_\delta (T)$, $m_{\rm ns}(T)$ and
$\Gamma_\delta (T)$ as calculated in Sections II and III.
We take a typical value $R=6$ fm for the transverse freeze-out radius of the
system. We define the thermal and chemical freeze-out temperatures
$T_{\rm th}$ and $T_{\rm ch}$ respectively as
$\tau_{\rm ther}(T_{\rm th})=R$ and $\tau_{\rm chem}(T_{\rm ch})=R$.
One finds from Fig.\ 2
\begin{equation}
T_{\rm th}\simeq 139 \;{\rm MeV }\quad {\rm and} \quad
T_{\rm ch}\simeq 168 \;{\rm MeV }\; ,\label{21}
\end{equation}
 which are the temperatures at which the thermal and
chemical equilibria start to break up, respectively.
It is worth noting that $T_{\rm th}$ is comparable to the
decoupling temperature of the thermal pions.

The result that $T_{\rm ch}$ is considerably higher than $T_{\rm th}$ offers an
interesting possibility to detect the suppression of the $U(1)_A$ anomaly
effect at high temperatures caused by the Debye-type screening. At
sufficiently high temperatures $T>T_{\rm ch}$, the $\eta$ rescattering
and the $\pi$-$\eta$ conversion are frequent so that the system possesses
both thermal and chemical equilibria. As the system expands and the
temperature falls into the range $T_{\rm th}<T<T_{\rm ch}$, the $\pi$-$\eta$
conversion process becomes slow and effectively is turned off; the system can
no longer maintain the chemical equilibrium.
There is an approximate conservation
of the total number of $\eta$'s since neither $\eta\eta\rightarrow \eta\eta$
or $\pi\eta\rightarrow \pi\eta$ can change the total $\eta$-number.
The number density of $\eta$ at the chemical break-up temperature
$T=T_{\rm ch}$ is determined by the mass of $\eta$ at such a temperature
$m_\eta (T_{\rm ch})$:
\begin{equation}
n_\eta \left[ m_\eta (T_{\rm ch}), T_{\rm ch}\right] =\frac{1}{2\pi^2}
m_\eta (T_{\rm ch})^2 T_{\rm ch}K_2\left[
\frac{m_\eta (T_{\rm ch})}{T_{\rm ch}}\right]\; ,
\label{22}
\end{equation}
and the momentum distribution is just the Boltzmann distribution with zero
chemical potential. However, this is not the final particle distribution,
because the thermal collisions can still alter the momentum distribution.
Nevertheless, the total
number $N_\eta$ given by
\begin{equation}
N_\eta =\pi R^2\tau_c n_\eta [m_\eta (T_{\rm ch}), T_{\rm ch}] \label{23}
\end{equation}
is conserved at any time $\tau <\tau_c$ since
$\eta\eta\leftrightarrow \pi\pi$ is turned off.
Here $m_\eta (T_{\rm ch})\simeq 360$ MeV and
$\tau_c$ is the proper time when the temperature of the
system reaches $T_{\rm ch}$.

As the system cools down to $T_{\rm th}$, the mass of $\eta$ should tend
to $m_\eta (T_{\rm th})\simeq 413$ MeV, according to Fig.~1.
If the rate for increasing $m_\eta (T)$ is
comparable to the thermal collision rate,
the $\eta$ particle adiabatically relaxes to $m_\eta (T_{\rm th})$.
In this case, which we shall call Scenario A, one expects a standard thermal
distribution for $\eta$ at the freeze-out temperature $T_{\rm th}$ with
a mass $m_\eta (T_{\rm th})$.
The total number conservation requires $\eta$ to
develop a chemical potential $\mu >0$ such that (neglecting
the transverse expansion)
\begin{equation}
\tau_de^{\mu/ T_{\rm th}}n_\eta \left[ m_\eta (T_{\rm th}),
T_{\rm th}\right] =
\tau_cn_\eta \left[ m_\eta (T_{\rm ch}) ,T_{\rm ch}\right] \; ,\label{24}
\end{equation}
where $\tau_d$ is the freeze-out time when $T=T_{\rm th}$.
The momentum distribution function in the local comoving frame is
\begin{equation}
f({\bf p})=e^{\mu/T_{\rm th}}e^{-\frac{\sqrt{m_\eta^2(T_{\rm th})
+{\bf p}^2}}{T_{\rm th}}}\; .
\label{25}
\end{equation}

The chemical potential $\mu$ is a function of temperature, whose value
at freeze-out can be determined from Eq.~(\ref{24}) once
$\tau_d/\tau_c$ is known. We assume that pions dominate the
energy-momentum tensor (in fact we explicitly checked the contribution
from $\eta$ and found it negligible) so that $\tau_d/\tau_c$
can be estimated by solving the ideal $1+1$ dimensional hydrodynamic equation
\begin{equation}
\frac{d\epsilon}{d\tau}+\frac{\epsilon +P}{\tau}=0 \: ,
\end{equation}
where $\epsilon$ is the energy density and $P$ is the pressure, for
massive pions. We find that $\tau_d/\tau_c \simeq 1.53$, given
$T_{\rm ch}/T_{\rm th}=1.21$. Substituting the ratio back in Eq.~(\ref{24}),
one finds $\lambda_\eta =e^{\mu/T_{\rm th}}=1.58$.  We thus predict
that if there is a partial $U(1)_A$ restoration at high temperatures,
the thermal $\eta$ production given by Eq.~(\ref{25}) will be enhanced
in this scenario due to both the finite chemical potential
$\lambda_\eta \simeq 1.58$ and a smaller $\eta$ mass
$m_\eta(T_{\rm th})\simeq 413$ MeV at the thermal freeze-out
temperature $T_{\rm th}$. To quantify such an
enhancement, we use Eq.~(\ref{25}) to calculate the $p_t$
distribution of $\eta$ particle, employing the fireball model and
taking into account the transverse flow effects
as described in Ref.\ \cite{hein}:
\begin{equation}
\frac{dN_\eta}{p_tdp_t}\propto \lambda_\eta
\int ^R_0rdr\sqrt{m_\eta^2+p_t^2}I_0({p_t\sinh \rho}/{T_{\rm th}})
K_1(\sqrt{m_\eta^2+p_t^2}\cosh\rho / T_{\rm th})\; ,\label{27}
\end{equation}
where $\rho ={\rm tanh}^{-1}(\beta_t)$ and $\beta_t=\beta_s(r/R)^\alpha$
(with $\beta_s=0.5$, $\alpha =2$) is the transverse flow
velocity profile \cite{hein}. To reduce the possible normalization ambiguity,
we also calculate the $p_t$-distribution for pions at the same freeze-out
temperature $T_{\rm th}$,  taking into account only the dominant
resonance decays, $\rho\rightarrow 2\pi$, and plot the ratio
\begin{equation}
\frac{\eta}{\pi^0}\equiv \frac{dN_\eta /p_tdp_t}{dN_\pi /p_tdp_t}
\label{28}
\end{equation}
as a function of $p_t$ in Figure 3. It should be noted that
the thermal ratio is only relevant when $p_t$ is small.
At very large $p_t$, hard processes become important and the fireball
model is no longer applicable.
For comparison, we also plot the same ratio for a normal case
in which the $\eta$ particles freeze out at the same temperature
$T_{\rm th}$ but with the zero-temperature mass $m_\eta=540$ MeV.

Another situation, which we shall call Scenario B, is that the rate
for increasing $m_\eta (T)$ when $T_{\rm th}<T<T_{\rm ch}$ is
considerably smaller than the thermal collision
rate. In this case, things get more complicated because
the screening process is out of
equilibrium. The $\eta$ number
conservation still holds, but the momentum distribution is quite
different from that in Scenario A. Roughly one may imagine that
even though the temperature drops to  $T_{\rm th}$ after
the chemical breakup,  $m_\eta$ will still have the
value $m_\eta (T_{\rm ch})$, in close analogy to a ``quenching''
situation. The number density at the thermal freeze-out
temperature is then $n_\eta [m_\eta (T_{\rm ch}),T_{\rm th}]$,
and the chemical potential is determined by
\begin{equation}
\tau_de^{\mu/ T_{\rm th}}n_\eta \left[ m_\eta (T_{\rm ch}) ,T_{\rm th}\right]
=\tau_cn_\eta \left[ m_\eta (T_{\rm ch}) ,T_{\rm ch}\right] \; , \label{29}
\end{equation}
yielding $\lambda_\eta =e^{\mu/ T_{\rm th}}\simeq 1.24$.
The momentum distribution function is
\begin{equation}
f({\bf p})=e^{\mu/T_{\rm th}}e^{-\frac{\sqrt{m_\eta^2(T_{\rm ch})
+{\bf p}^2}}{T_{\rm th}}}\; ,
\label{30}
\end{equation}
which predicts larger $\eta$ enhancement at low $p_t$ than at high $p_t$.
We also plot the ratio $\eta/ \pi^0$ based on this scenario in Fig.~3.

What happens after the thermal freeze-out? It is clear that there must
exist some mechanism for the $\eta$ to relax from the `temporary' entity
whose mass is either $m_\eta (T_{\rm th})$ or $m_\eta (T_{\rm ch})$ to
its true identity at zero temperature with $m_\eta =540$ MeV. A possible
picture might be that the $\eta$ particles still
feel a negative potential in the fireball. The height of the potential
barrier is determined by the mass difference
$\Delta m=m_\eta -m_\eta (T_{\rm ch})$.  The $\eta$ particles
with $p_t$ smaller than $\Delta m$ will be trapped in the potential
well until the rarefaction wave reaches the center
of the interaction volume. Such a picture has been suggested by Shuryak
\cite{shur} and is similar to the mechanism of cold kaon
production \cite{koch}.  At this stage,
we do not attempt to address this nonequilibrium issue, but just to
remark that our calculation here may have underestimated the
enhancement effect at small
$p_t$\raisebox{-.5ex}{$\stackrel{<}{\sim}$}$\Delta m \sim 100 - 200$ MeV.

Both Scenarios A and B predict an enhancement of the thermal $\eta$
production in the light of a partial $U(1)_A$ symmetry
restoration. It would be very interesting to test the idea experimentally
by measuring the ratio $\eta/\pi^0$, especially its $p_t$ dependence.
Although preliminary data from WA80 \cite{wa80} on $\eta/\pi^0$
ratio in both central and peripheral $S+Au$ collisions at
the CERN SPS energy, as indicated in Fig.~3, have shown
such a trend of enhancement, one certainly needs better
statistics in order to make a definite conclusion.
A related matter is the enhanced
dilepton pair production via $\eta$ Dalitz decay $\eta\rightarrow
\ell^+\ell^-\gamma$. If the $\eta$ production
is enhanced about 3 times,  as we have
predicted, the observed dilepton enhancement with the invariant mass
below 500 MeV at the CERN SPS \cite{sps} may be partially
accounted for.

\section{Role of the $\eta '$}
\label{V}
There should be also some enhancement of the ratio $\eta '/\pi^0$, since
the mass of $\eta '$ also decreases as the temperature increases.
Moreover, since the couplings of $\eta '$ to $\eta$ and $\pi$
in our model become small and eventually goes to zero when $U(1)_A$
is completely restored, $\eta '$ might decouple from the system
earlier than $\eta$.
The decay $\eta '\rightarrow \pi\pi\eta$ can also enhance
the $\eta$-production.
However, in our model, we postulate that the $U(1)_A$ restoration occurs
at a temperature below the chiral phase transition temperature.
Therefore, the kaon mass $m_K$ is large and  the $\eta '$
does not become very soft. At $T=T_{\rm ch}$, the $\eta '$
mass $m_{\eta '}$ is about 750 MeV.
Even without the effect of chiral symmetry breaking, the large
strange-quark mass can give rise to a large mixing between $\eta$
and $\eta '$ according to Eq.~(\ref{4}). This mixing gives $\eta '$
a mass $m_{\eta '}^2=2m_K^2-m_\pi^2$ even if $U(1)_A$ is
completely restored. This mass is significantly larger
than the $\eta$ mass at any temperature.  Therefore, in the
context of our model,  the $\eta '$ effects are only
moderate. The ratio           $\eta '/\pi^0$ should never
exceed that of $\eta/\pi^0$.

\section{Role of QCD Sphalerons}
\label{VI}
So far we have confined ourselves to the possible suppression
of the instanton effects at finite temperature that causes the
softening of the masses arising from the topological charge
transitions. At very high temperatures, it is known that such
transitions can occur without going through the instanton
configurations. In fact, they are dominated by
sphaleron-like transitions whose electroweak counterparts
have been extensively studied in the literature \cite{sph}.
It is pointed out by McLerran, Mottola and Shaposhnikov \cite{qcdsp}
that the rate of a QCD sphaleron transition should be estimated in analogy
to the electroweak theory for temperatures above the symmetry-restoration,
which may not be quite suppressed.
In the range of temperatures discussed in this paper,
the rate of the QCD sphaleron transition may be unimportant. A rough
estimate by Giudice and Shaposhnikov \cite{qcdsp2} is
\begin{equation}
\Gamma_{\rm sph}^{\rm QCD}=\frac{8}{3}(\alpha_s/\alpha_W)^4
\Gamma_{\rm sph}^{\rm EW}=\frac{8\kappa}{3}(\alpha_sT)^4\; ,
\end{equation}
where $\kappa$ is the strength of the transition.
The characteristic time scale of the
sphaleron transition is
\begin{equation}
\tau_{\rm sph}=(192\kappa\alpha_s^4T)^{-1}\sim \frac{50}{\kappa T}\; .
\end{equation}
There is some evidence for $\kappa$ to be ${\cal O}(1)$
from lattice calculations \cite{qcdsp2}.
Unless $\kappa$ is really big, greater than 10,
the sphalerons should be decoupled from the system
in the hadronic phase, where the instanton effect is most dominant.

\section*{Acknowledgements}
We wish to thank M.\ Suzuki and V.\ Koch
for helpful discussions.
This work was supported by the Director, Office of Energy
Research, Office of High Energy and Nuclear Physics, Divisions of High
Energy Physics and Nuclear Physics of the U.S. Department
of Energy under Contract No.\
DE-AC03-76SF00098, and by the
 Natural Sciences and Engineering
Research Council of Canada.

{\it Note added:}  After completing this work we learned of a
recent paper by J.~Kapusta, D.~Kharzeev and L. McLerran on the
effect of $U(1)_A$ symmetry restoration on $\eta '$ particle
production \cite{kkm}.

\newpage
\centerline{\bf Figure Captions}
\vskip 15pt
\begin{description}
\item[Fig. 1] The temperature dependence of $m_\eta$, $m_{\eta '}$,
$m_\delta$. The parameter in the exponential suppression of the
instanton effect is taken to be $T_0=150$ MeV.
\item[Fig. 2] The characteristic time scales of the thermal and
chemical equilibration for the $\eta$ particle.
\item[Fig. 3] The predicted ratio $\eta/\pi^0$ as a function of
the transverse momentum $p_t$ in three scenarios as discussed in the text.
The preliminary data from WA80 \cite{wa80} are also indicated.
\end{description}
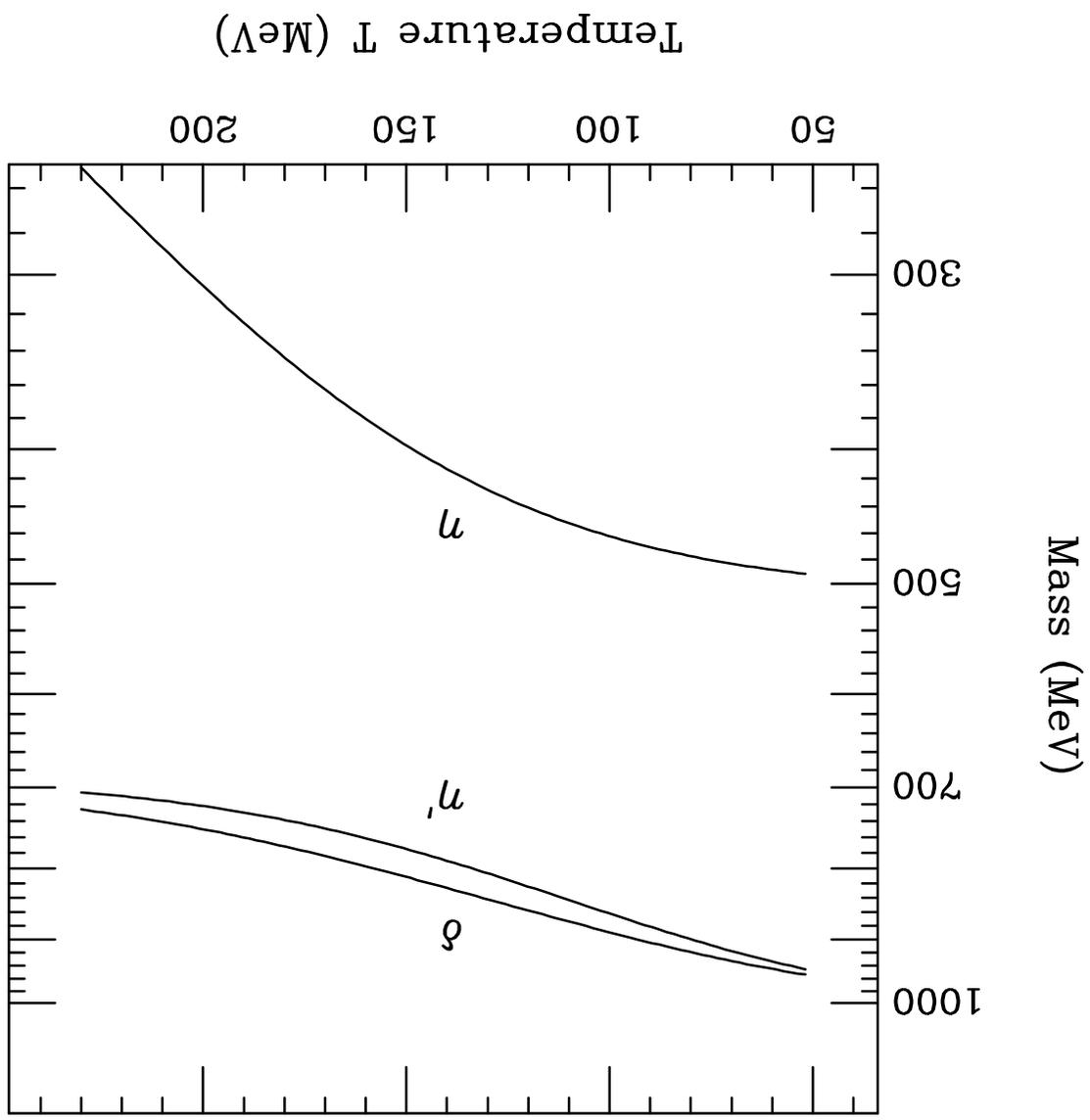
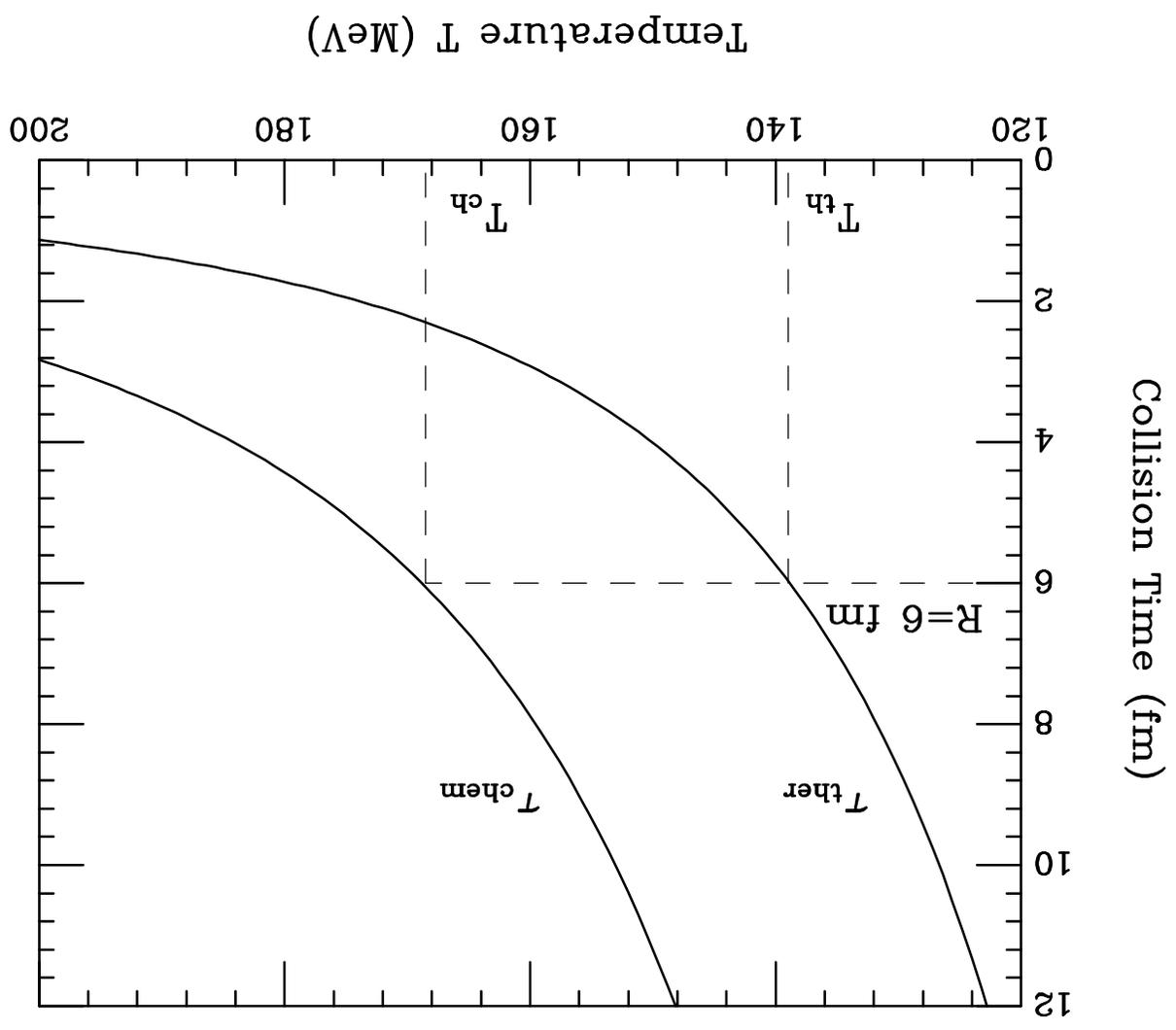
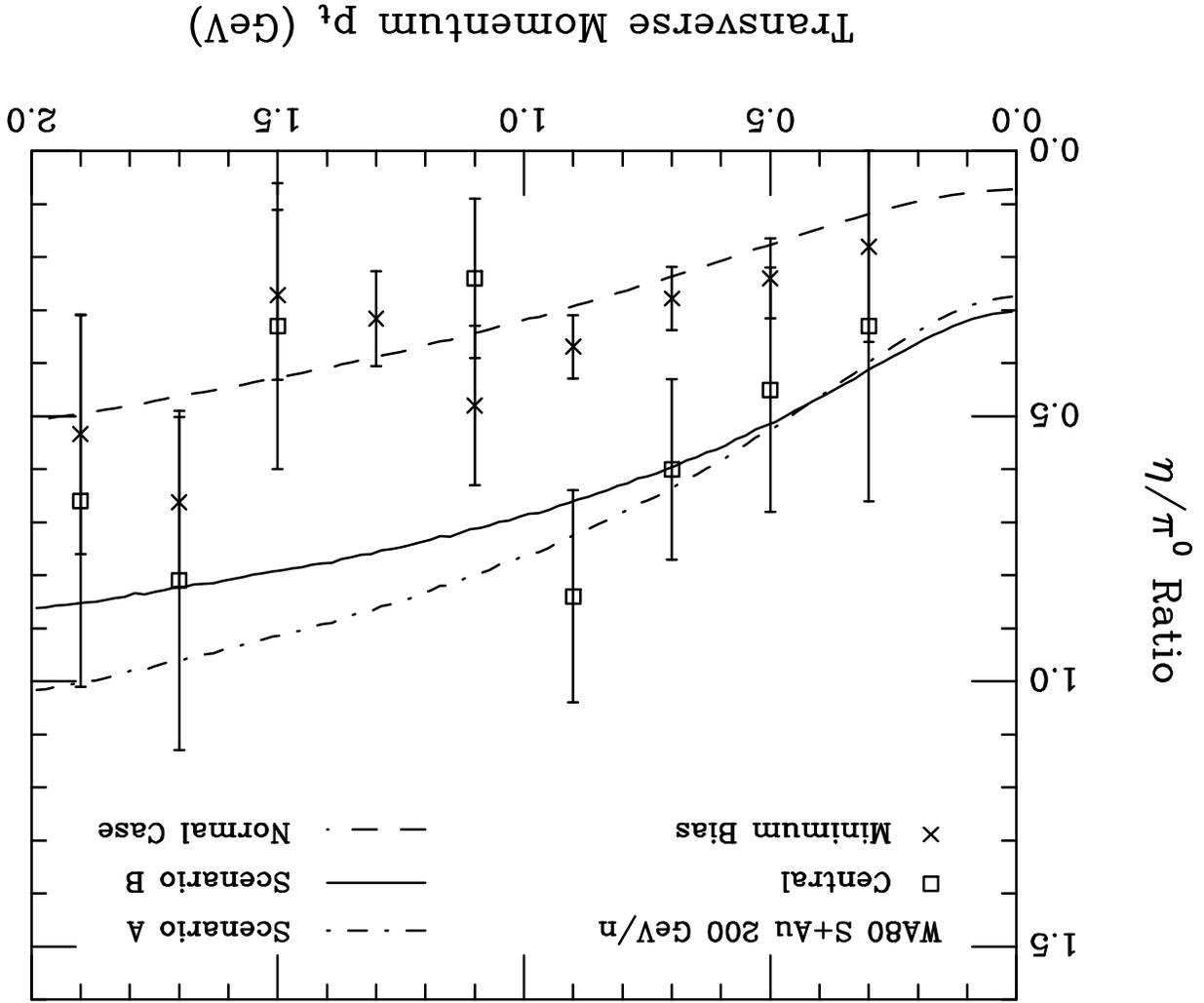
\end{document}